\documentclass[conference]{IEEEtran}
\IEEEoverridecommandlockouts
\usepackage{cite}
\usepackage{amsmath,amssymb,amsfonts}
\usepackage{algorithmic}
\usepackage{graphicx}
\usepackage{textcomp}
\def\BibTeX{{\rm B\kern-.05em{\sc i\kern-.025em b}\kern-.08em
    T\kern-.1667em\lower.7ex\hbox{E}\kern-.125emX}}

\begin{document}

\title{Performance of Source transmit Antenna selection for MIMO cooperative communication System Based DF protocol: Symbol Error Rate and Diversity order}

\author{\IEEEauthorblockN{Mokhtar Bouteggui}
\IEEEauthorblockA{\textit{USTHB University} \\
\textit{Telecommunications Department,LISIC Laboratory}\\
Algiers, Algeria \\
bouteguimokhtar@gmail.com}
\and
\IEEEauthorblockN{Fatiha Merazka}
\IEEEauthorblockA{\textit{USTHB University} \\
\textit{Telecommunications Department,LISIC Laboratory}\\
Algiers, Algeria \\
fmerazka@usthb.dz}}
\maketitle

\begin{abstract}
In this work, we study the performance of a single relay Multiple Input Multiple Output (MIMO) cooperative communication system based on Decode and Forward (DF) relaying protocol for two strategies using transmit antenna selection at the source. The first strategy uses one antenna between the relay and the destination, and the second strategy uses Space Time Block Coding (STBC). All channels follow the Rayleigh fading distribution. We derive the expression and Upper Bound for Symbol Error Rate (SER) for M-ary Phase Shift Keying (M-PSK), and the diversity order for both strategies. The analytical results show that the second strategy performs better than the first one for the same diversity order and the same Rate R.
\end{abstract}


\section{Introduction}
Multiple-input-multiple-output (MIMO) systems use more than one antenna at the transmitter and/or the receiver which provides better capacity and reliability than Single-input-single-output SISO systems \cite{b1}\cite{b2}. In many situations, however, due to limitation on size, processing power, and cost, it is not practical for some users, especially small wireless mobile devices, to be equipped with multiple antennas.

In order to cope with this problem, cooperative communication attract continuous attention these years. With cooperative communication, antennas of different nodes can form a virtual MIMO system. By exploiting the spatial diversity provided by this virtual MIMO system, communication performance can be improved \cite{b3}\cite{b4}.

The two famous relaying protocols are Amplify and Forward (AF) and DF \cite{b5}. In AF, the relay simply amplifies its received signal from the transmitter based on its power constraint and forwards to the receiver. In DF, each relay or relay antenna conducts hard decoding of the information based on the signal it receives, and then forwards its decoded information to the receiver either directly or after re-encoding. The combination of MIMO and relay technologies is often referred to as the MIMO relay technology, where multiple antennas are mounted on the source, relay, and destination nodes in order to achieve both multiplexing and diversity gains.

Several works in existing literature have analyzed the performance of DF MIMO cooperative communication. In \cite{b6} the authors analyze the performance of single and multiple relay MIMO STBC system and derive the end to end closed form expression diversity order and optimal power allocation.

In \cite{b7} the authors combine both Maximum Ratio Combining (MRC) and Distributed Space-Time Coding (DSTC) for multiple antenna nodes to enhance the performance of the original DSTC. Both systems in \cite{b6} and \cite{b7} do not require Channel State Information (CSI) at the transmission side but only at the receiver. When CSI is available at the source, transmit antenna selection is usually applied to MIMO system to solve the problem of radio frequency chain which are expensive in terms of size, power and hardware. The authors in \cite{b8} analyze the performance of two strategies for cooperative MIMO system with single relay, in the first strategy which maximizes the Signal to Noise Ratio (SNR) of the source-destination and relay-destination channel which have diversity order $N_SN_D+N_R$, where $N_S,N_R$ and $N_D$ are the number of antenna at the Source, Relay and Destination, respectively. In the second strategy which maximizes the SNR of source-relay and relay-destination which achieves diversity order $N_D+N_R\min\{N_S,N_D \}$.

In \cite{b9} the authors propose source transmit antenna selection strategy for one source, one destination and $N$ relays with $ N_S,N_D$ and $N_R$ antennas respectively which selects $N_{ss}$ antenna among $N_s$ antenna at the source which achieves the maximum diversity order $N_SN_D+NN_R \min\{N_S,N_D\}$. In the case, when the relay does not have the amplitude of the MIMO channel, in other words, the CSI is not available at the relay, various transmission schemes can be used at the relay phase.  For example, we can select one antenna randomly at the transmitter and receiver to transmit and receive simultaneously, also transmit and receive diversity can be used in the relay-destination links such as STBC in the case of Multiple Input Single Output (MISO) or MRC in the case of Single Input Multiple Output (SIMO), Spatial Multiplexing (SM) also can be used for higher speed.

In this paper, we analyze the performance of a single relay based on selective DF cooperative MIMO system with $M-PSK$ by deriving the Symbol Error Rate (SER) expression and upper bound, also the diversity order for two strategies. The first strategy uses a single transmit and receive antenna in relay phase and the second strategy uses transmit diversity using STBC between the relay and destination using the same transmit antenna selection strategy at the source in \cite{b8}.

Throughout this paper, the following notation is employed $(\cdot)^H,(\cdot)^*, \|\cdot\|_F, Tr(\cdot)$ and $E\{∙\cdot\}$ which denote the Hermitian transpose, complex conjugate, Frobenious norm, trace and expectation, respectively, $|\cdot|$ denotes the absolute value or the cardinality of a set.

\section{System Model}
Consider a wireless cooperative MIMO system with one source, one relay and one destination as shown in Fig. 1. The source node $S$, relay node $R$, and destination node $D$ are equipped with $N_S, N_R$, and $N_D$ antennas, respectively. The relay uses the DF protocol and forward to destination only when decode correctly. Denote the channel matrix from the source to the relay as $H_{SR}\in C^{N_R×N_S}$ the and channel matrix from the source to the destination as $H_{SD}\in C^{N_d×N_S}$ and channel matrix from the relay to the destination as $H_{RD}\in C^{N_d×N_s}$ with coherence interval $T_1$.
%%%%%%%%%%%%%
\begin{figure}[htbp]
\centerline{\includegraphics[scale=0.4]{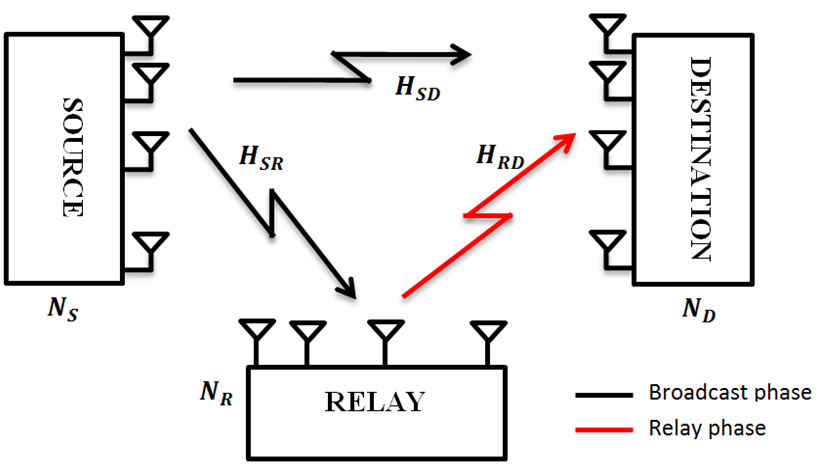}}
\caption{DF MIMO relay network.}
\label{fig1}
\end{figure}
%%%%%%%%%%%%%%
Let $h_{d_i s_j}\in H_{SD},h_{r_i s_j }\in H_{SR}$ and $h_{d_i r_j}\in H_{RD}$ the channel coefficients from the $i^{th}$ receiver antenna to the $j^{th}$ transmitter antenna in the source-destination, source-relay and relay-destination channel respectively.

We assume that the channels coefficients are modeled as zero-mean, complex Gaussian random variables with unit variance channels are independent, the magnitude is Rayleigh distributed.

Let $g_{d_i s_j}=|h_{d_i s_j}|^2,g_{r_i s_j }=|h_{r_i s_j} |^2$ and $g_{d_i r_j}=|h_(d_i r_j|^2$, the channels gains from the $i^{th}$ receiver antenna to the $j^{th}$ transmitter antenna in the source-destination, source-relay and relay-destination channel, which are exponentially distributed with the parameter $1/(\sigma_{sd}^2), 1/(\sigma_{sr}^2)$ and $1/(\sigma_{rd}^2)$ respectively.

The transmission can be split into two phases, a broadcast phase and a relay phase. In the broadcast phase, the source takes $T_1$ time slot to broadcasts data symbols to the relay and the destination simultaneously using the Time Division Multiplexing Access (TDMA) scheme. In the relay phase, the relay decodes and forwards the received data symbols to the destination using Space Time Coding (STC) scheme and take $T_2$ time slot.

\subsection{The Broadcast Phase}
The symbols are transmitted into a vector $\mathbf{x}\in \mathbb{C}^{T_1\times1}$, with the normalization $E\{\mathbf{x}^{*} \mathbf{x}\}=1$.
Let $y_{SD}\in \mathbb{C}^{T_1\times1},y_{SR} \in \mathbb{C}^{T_1\times1}$ the received vectors at the destination and relay respectively. So we can write
\begin{equation}
y_{SD}=\sqrt{P_sT_1}h_{D_i S_j}\times x+w_{SD}\label{eq1}
\end{equation}
\begin{equation}
y_{SR}=\sqrt{P_sT_1}h_{R_i S_j}\times x+w_{SR}\label{eq2}
\end{equation}
where $w_{SD}\in \mathbb{C}^{T_1\times 1},w_{SD}\in \mathbb{C}^{T_1\times 1}$ noise vector between source-destination and source-relay. The noise vectors are modeled as, zero-mean, circularly symmetric complex Gaussian random variable with variance $N_0$.
Let $y_{SD}^k$ the received symbol at the $k^{th}$ time slot as follows:
\begin{equation}
y_{SD}^k=\sqrt{P_s}h_{{D_i}{S_j}}\times x^k+w_{SR}^k\label{eq3}
\end{equation}
We define $\alpha_{SD}$ the weight factor of S-D link in \eqref{eq4} such that the SNR is maximizing as:
\begin{equation}
\alpha_{SD}=h_{D_i S_j}^*\label{eq4}
\end{equation}
Then for the received SNR is:
\begin{equation}
  	\lambda_{SD}=\frac{P_s}{N_0}|h_{D_i S_j}|^2\label{eq5}
\end{equation}
Similarly for the link S-R:
\begin{equation}
 	\alpha_{SR}=h_{R_i S_j}^*\label{eq6}
\end{equation}
\begin{equation}
	\lambda_{SR}=\frac{P_s}{N_0}|h_{R_i S_j}|^2\label{eq7}
\end{equation}
\subsection{The Relay Phase}
\paragraph{Strategies I-Single Transmit and receive antenna between the relay and destination}

In this phase, the relay chooses one antenna randomly to transmit and the receiver chooses one antenna randomly to receive as show in Fig. 2.

Lets $y_{RD}^{k+T_1}$ the received data flow $x^k$ at the destination at the $k+T_1$ time slot during the relay phase which can be written as:
\begin{equation}
y_{RD}^{k+T_1}=\sqrt{P_R}h_{D_i S_j}\times x^k+w_{RD}^k\label{eq8}
\end{equation}

Similarly we define $\alpha_{RD}$ the weight factor of R-D link in Equation \eqref{eq9} such that the SNR is maximizing as:
\begin{equation}
	\alpha_{RD}=h_{D_i R_j}^*\label{eq9}
\end{equation}
Then for the received SNR is
\begin{equation}
	\lambda_{RD}=\frac{P_R}{N_0}|h_{D_i R_j}|^2\label{eq10}
\end{equation}
The SNR of the combined signal at the MRC detector can be written as \cite{b1}:
\begin{equation}\label{eq11}
  \lambda=\frac{P_s|h_{D_i S_j}|^2+P_R |h_{D_i R_j}|^2}{N_0}
\end{equation}	
%%%%%%%%%%%
\begin{figure}[htbp]
\centerline{\includegraphics[scale=0.4]{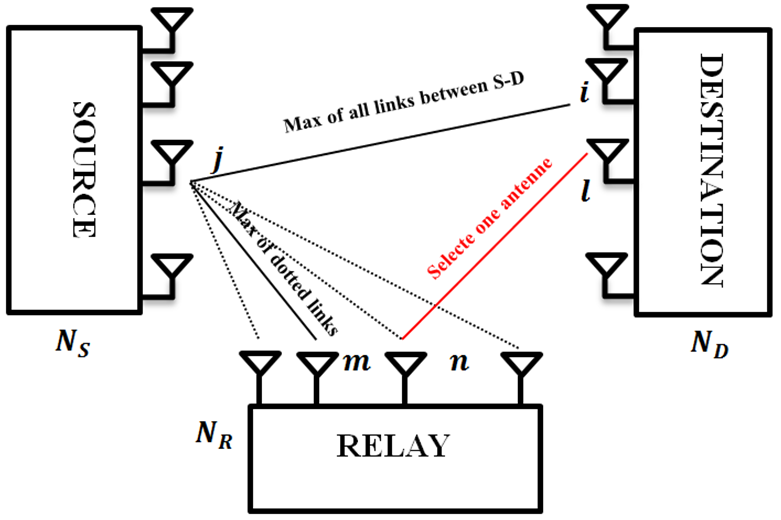}}
\caption{Step of the first Strategy: single transmit and receive antenna in the relay phase..}
\label{fig2}
\end{figure}
%%%%%%%%%%%
\paragraph{Strategy II-Space Time Block coding between the relay and destination}
The broadcast phase is similar to the previously strategy, after receiving the transmitted vector $\mathbf{x}$, the relay generates the transmitted codeword matrix  $X\in \mathbb{C}^{N_R\times T_2}$ with $N_R$ transmitted antennas and $T_2$ time slot from the relay to the destination from the codeword matrix set $\mathcal{C}$ . The received codeword $Y_{RD}\in \mathbb{C}^{N_D\times T_2}$ at destination is \cite{b1}:

\begin{equation}\label{eq12}
Y_{RD}=\sqrt{ \frac{P_r}{N_R}}H_{RD}\times X+W_{RD}
\end{equation}

where $H_{RD}\in \mathbb{C}^{N_D\times N_r}$ the channel matrix from the relay to the destination, and $W_{RD}$ the noise matrix between the relay and destination.
Since we assume that the perfect channel information is available at the receiver side and the noise components are independent, then the Maximum-Likelihood (ML) decoding of the transmitted matrix is \cite{b1}:

\begin{equation}\label{eq13}
\hat{X}=\arg⁡\max_{X\in\mathcal{C}}
\begin{Vmatrix}
  ⁡ Y_{RD}-\sqrt{\frac{P_r}{N_R}}H_{RD}X
\end{Vmatrix}_{F}^2
\end{equation}

\section{SER analysis of antenna selection strategies at the source}
\subsection{SER of Strategies I}
In the broadcast phase which involves 2 steps, in the first step the $i^{th}$ receive antenna and the  $j^{th}$ transmit antenna are selected based on the maximum channel gain $g_{SD}$ of all the source destination channels. In the second step the $k^{th}$ antenna at the relay are selected based on the maximum channel gain $g_{SR}$ between the $j^{th}$ transmit antenna at the source and all channels at the relay. In the relay, the $n^{th}$ antenna at the destination and the $m^{th}$ antenna at the relay are selected randomly. We can write \cite{b8}:
\begin{align}
g_{SD} &= \max\{g_{D_i S_j}\},\text{ }   1\leq D_i \leq N_D,1 \leq S_j \leq N_S, \nonumber\\
g_{SR} &= \max\{g_{R_i S_j}\},\text{ }   1\leq R_i \leq N_R,                     \nonumber\\
g_{RD} &= g_{D_i R_j},\text{ } i\in [1,2,\cdots,N_D],j\in[1,2,\cdots,N_R].
\label{eq14}
\end{align}
The probability density functions (PDF) are:
\begin{align}
f_{g_{SD}}(g) &= \frac{N_SN_D}{\sigma_{SD}^2}
exp\begin{pmatrix}
\frac{-g}{\sigma_{SD}^2}
\end{pmatrix}
\begin{pmatrix}
1-exp\begin{pmatrix}
\frac{-g}{\sigma_{SD}^2}
\end{pmatrix}
\end{pmatrix}^{N_SN_D-1},\nonumber\\
f_{g_{SR}}(g)&=\frac{N_R}{\sigma_{SR}^2}
exp
\begin{pmatrix}
\frac{-g}{\sigma_{SR}^2}
\end{pmatrix}
\begin{pmatrix}
1-exp\begin{pmatrix}
\frac{-g}{\sigma_{SR}^2}
\end{pmatrix}
\end{pmatrix}^{N_R-1}, \nonumber\\
 f_{g_{RD}}(g) &=\frac{1}{\sigma_{RD}^2}
exp\begin{pmatrix}
\frac{-g}{\sigma_{RD}^2}
\end{pmatrix}.
\label{eq15}
\end{align}
The End-to-End SER for the system can be written as \cite{b1}:
\begin{equation}\label{16}
P_e=P_{S\rightarrow D}\times P_{S\rightarrow R}+P_{S\rightarrow D,R\rightarrow D}\times (1-P_{S\rightarrow R})
\end{equation}

Let’s define the probability of symbol error for M-PSK as follow \cite{b8}:
\begin{equation}\label{17}
\Psi(\lambda_{SD})=\frac{1}{\pi}\int_{0}^{(\frac{M-1}{M})\pi}exp\begin{pmatrix}
  -\frac{b}{\sin^2\theta}\lambda_{SD}
\end{pmatrix}d\theta
\end{equation}
and the Gaussian Q-function as \cite{b6}:
\begin{equation}\label{18}
Q(x)=\frac{1}{\pi}\int_{0}^{\frac{\pi}{2}}exp\begin{pmatrix}
  -\frac{x^2}{2\sin^2\theta}
\end{pmatrix}d\theta
\end{equation}
where $b=\sin^2 \begin{pmatrix}\frac{\pi}{M}\end{pmatrix}$. For M-PSK, the error probability for the link Source-Destination is \cite{b8}:
\begin{align}\label{19}
P_{S\rightarrow D}&=E_{g_{SD}}\{\Psi(\lambda_{SD})\}\nonumber\\
& = N_SN_d\sum\limits_{n=0}^{N_SN_d-1}\displaystyle\binom{N_SN_d-1}{n}(-1)^n\nonumber\\
&F\left(\dfrac{bP_S\sigma_{SD}^2}{N_0\sin^2\theta}+n+1\right)
\end{align}
where $F(x(\theta))=\int_{0}^{(\frac{M-1}{M})\pi}\frac{1}{x(\theta)}d\theta$, (see Appendix for (19)). Similarly the probability for the link Source-Relay is:
\begin{align}\label{20}
P_{S\rightarrow R}&=E_{g_{SR}}\{\Psi(\lambda_{SR})\}\nonumber\\
& = N_R\sum\limits_{n=0}^{N_R-1}\displaystyle\binom{N_R-1}{n}(-1)^n\nonumber\\
&F\left(\dfrac{bP_S\sigma_{SR}^2}{N_0\sin^2\theta}+n+1\right)
\end{align}
The error probability at the destination when the relay decodes correctly during the broadcast phase is (see Appendix for (21)) :
\begin{align}\label{21}
P_{S\rightarrow D, R\rightarrow D}&=E_{{f_{g_{SD}}}f_{g_{RD}}}\{\Psi(\lambda)\}\nonumber \\
&=N_{S}N_d\sum\limits_{n=0}^{N_SN_d-1}\displaystyle\binom{N_SN_d-1}{n}(-1)^n\nonumber\\
&F\left(\left(\dfrac{bP_S\delta_{SD}^2}{N_0\sin^2\theta}+n+1\right)\left(\dfrac{bP_R\delta_{RD}^2}{N_0\sin^2\theta}+1\right)\right)
  \end{align}
So, substituting in \eqref{16}, we get the expression of SER for M-PSK.
\paragraph{SER upper bound and Diversity order}
Replacing $1-P_{S\rightarrow R}\approx1$, we can get the following upper bound as (see Appendix for (22)):
\begin{align}\label{22}
   P_e\leq &\dfrac{N_{S}N_dN_R}{\pi^2}\dfrac{(N_r-1)!}{\left(\dfrac{bP_S\delta_{SR}^2}{N_0}\right)^{N_r}}\dfrac{(N_SN_d-1)!I_1I_2}{\left(\dfrac{bP_S\delta_{SD}^2}{N_0}\right)^{N_SN_d}}
\nonumber \\
&+\dfrac{N_{S}N_d}{\pi}\dfrac{(N_SN_d-1)!}{\left(\dfrac{bP_S\delta_{SD}^2}{N_0}\right)^{N_SN_d}}\dfrac{I_3}{\left(\dfrac{bP_R\delta_{RD}^2}{N_0}\right)}
\end{align}
where
\begin{align}
  I_1 &= \int_{0}^{(\frac{M-1}{M})\pi}(\sin\theta)^{2N_R}d\theta    \\
  I_2 &= \int_{0}^{(\frac{M-1}{M})\pi}(\sin\theta)^{2N_SN_D}d\theta \\
  I_3 &  \int_{0}^{(\frac{M-1}{M})\pi}(\sin\theta)^{2N_SN_D+1}d\theta
\end{align}
Subtitling $P_S=\beta_1 P$ and $P_R=\beta_2 P$ where $\beta_1+\beta_2=1$ in \eqref{22} we get the diversity order of the system as $N_s N_d+1$ which is the exponent of SNR in \eqref{26}:
\begin{align}\label{26}
P_e&\leq \frac{N_SN_DN_R}{\pi^2}\nonumber\\ &\frac{(N_SN_D-1)!}{(b\beta_1\sigma_{SR}^2)^{N_R}}\frac{(N_SN_D-1)!I_1I_2}{(b\beta_1\sigma_{SD}^2)^{N_SN_D}}\begin{pmatrix}
                                                                                                                                      \frac{N_0}{P}                                                                                                                                 \end{pmatrix}^{N_R+N_SN_D}\nonumber\\
&+\frac{N_SN_D}{\pi}\frac{(N_SN_D-1)!}{(b\beta_1\sigma_{SD}^2)^{N_SN_D}}\frac{I_3}{(b\beta_2\sigma_{RD}^2)}\begin{pmatrix}\frac{N_0}{P}
                                                                                                                                    \end{pmatrix}^{N_SN_D+1}
\end{align}
\paragraph{Optimal Power Allocation}

Substituting $P_R$ as $P-P_S$ in \eqref{26}, and differentiating it with respect to $P_S$ and equating it to zero we get:
\begin{align}\label{27}
\frac{\Delta_1(N_R+N_SN_D)P^{-(N_R+N_SN_D+1)}}{(\frac{P_S}{P})^{(N_R+N_sN_D+1)}}  & \nonumber\\
+\frac{\Delta_2(N_SN_D)P^{-(N_SN_D+1)}}{(\frac{P_S}{P})^{(N_SN_D+1)}(1-\frac{P_S}{P})} &\nonumber\\
-\frac{\Delta_2P^{-(N_SN_D+1)}}{(\frac{P_S}{P})^{(N_SN_D)}(1-\frac{P_S}{P})^2}=0&
\end{align}
where $\Delta_1$ and $\Delta_2$ are given in \eqref{28}. Finally, finding the roots of \eqref{27} which lie in the interval $[0,1]$, we get the optimal power allocation.
\begin{align}\label{28}
  \Delta_1 &=
  \frac{N_SN_DN_R(N_R-1)!(N_SN_D-1)!I_1I_2}{\pi^2(\frac{b\sigma_{SR}^2}{N_0})^{N_R}(\frac{b\sigma_{SD}^2}{N_0})^{N_SN_D}}
   \nonumber\\
  \Delta_2 &=\frac{N_SN_D(N_SN_D-1)!I_3}{\pi(\frac{b\sigma_{SD}^2}{N_0})^{N_SN_D}(\frac{b\sigma_{SD}^2}{N_0})}
\end{align}
%%%%%%%%%%%%%%%%%%%%%%%%%%%%%%%%%%%%%%%%%%%%%%%%%%%%%%%%%
\subsection{SER strategies II}
Similarly to the first strategy, the error probability for the links Source-Destination $P_{S\rightarrow D}$, and Source-Relay $P_{S\rightarrow R}$  are the same as \eqref{19} and \eqref{20}.
Let’s define the Pairwise Error Probability (PEP) as the probability when a transmitted codeword matrix $X_n$ mistaken by another codeword matrix $X_l$ at the receiver for a fixed channel during one codeword matrix as \cite{b1}:
\begin{align}\label{29}
  P(X_n\rightarrow X_l|H_{RD}) =& \nonumber \\
   &Q\begin{pmatrix}
        \sqrt{\frac{P_R}{2N_RN_0}H_{RD}\|X_n-X_l\|_F^2}
      \end{pmatrix}
\end{align}
Averaging the PEP over the over the channel distribution, the average PEP can be determined as follows \cite{b6}:
\begin{align}\label{30}
  P(X_n\rightarrow X_l) =&E_{H_{RD}}\{P(X_n\rightarrow X_l|H_{RD})\} \nonumber \\
   =&G\begin{pmatrix}
        \prod_{k=1}^{N_R}\begin{pmatrix}
                           1+\frac{P_R\lambda_{k,l}\sigma^2_{RD}}{4N_0N_R\sin^2\theta}
                         \end{pmatrix}^{N_D}
      \end{pmatrix}
\end{align}
where $\lambda_{1,l},\lambda_{2,l},\cdots,\lambda_{N_R,l}$, are the non-zero eigenvalues of $(X_n-X_l )(X_n-X_l )^H$, and $G(x(\theta))=\int_{0}^{\pi/2}\frac{1}{x(\theta)}d\theta$.
The error probability for the link Relay-Destination can be upper bounded using the union bound which is very tight at high SNR as \cite{b6}:
\begin{align}\label{31}
P_{R\rightarrow D} \leq& \sum_{X_l\in \mathcal{C},l\neq n}^{|\mathcal{C}|}P_{R\rightarrow D}(X_n\rightarrow X_l)\nonumber\\
=& \sum_{X_l\in \mathcal{C},l\neq n}^{|\mathcal{C}|}G\begin{pmatrix}
        \prod_{k=1}^{N_R}\begin{pmatrix}
                           1+\frac{P_R\lambda_{k,l}\sigma^2_{RD}}{4N_0N_R\sin^2\theta}
                         \end{pmatrix}^{N_D}
      \end{pmatrix}
      \end{align}
The error probability at the destination when the relay decodes correctly during the broadcast phase is (see Appendix for \eqref{32}):
\begin{align}\label{32}
P_{S\rightarrow D, R\rightarrow D}&=E_{f_{g_{SD}}}f_{g_{RD}}\{\Psi(\lambda)\}\nonumber \\
&=N_{S}N_d\sum\limits_{n=0}^{N_SN_d-1}\displaystyle\binom{N_SN_d-1}{n}(-1)^n\nonumber\\
&F\left(\dfrac{bP_S\delta_{SD}^2}{N_0\sin^2\theta}+n+1\right)\nonumber\\
&\sum\limits_{X_l\in \mathcal{C}, l\neq n}^{|C|}F\left(\Pi_{k=1}^{N_r}\left(\dfrac{P_r\lambda_{k,l}\delta_{rd}^2}{4N_0N_R\sin^2\theta}+1\right)^{N_d}\right)
\end{align}
\paragraph{SER upper bound and Diversity order}
Replacing $1-P_{S\rightarrow R}\approx 1$ and approximating at high SNR, we can get the following upper bound as:
\begin{align}\label{33}
  P_e \leq&\frac{N_SN_DN_R}{\pi^2}\frac{(N_R-1)!}{\begin{pmatrix}
                                           \frac{bP_S\sigma_{SR}^2}{N_0}
                                         \end{pmatrix}^{N_R}}\frac{(N_SN_D-1)!I_1I_2}{\begin{pmatrix}
                                           \frac{bP_S\sigma_{SD}^2}{N_0}
                                         \end{pmatrix}^{N_SN_D}}
   \nonumber\\
  +&\frac{N_SN_D}{\pi}\frac{(N_SN_D-1)!I_2}{\begin{pmatrix}
                                           \frac{bP_S\sigma_{SD}^2}{N_0}
                                         \end{pmatrix}^{N_SN_D}}\frac{I_4}{\begin{pmatrix}
                                           \frac{P_R\xi\sigma_{RD}^2}{N_0}
                                         \end{pmatrix}^{N_RN_D}}
\end{align}
where
\begin{align}
I_4 =& \int_{0}^{\pi/2}(\sin\theta)^{2N_RN_D}d\theta\\
\lambda_l =& \prod_{k=1}^{N_R}\lambda_{k,l} \\
\xi =& \sum_{X_l\in \mathcal{C},l\neq n}^{|\mathcal{C}|}\lambda_l.
\end{align}
Subtitling $P_S=\beta_1 P$ and $P_R=\beta_2 P$ where $\beta_1+\beta_2=1$, we get the diversity order of the system as $N_SN_D +N_R$ from \eqref{37}.
\begin{align}\label{37}
P_e\leq&\frac{N_SN_DN_R}{\pi^2} \frac{(N_R-1)!}{(b\beta_1\sigma_{SR}^2)^{N_R}}\frac{(N_SN_D-1)!I_1I_2}{(b\beta_1\sigma_{SD}^2)^{N_SN_D}}\nonumber\\
&\times\begin{pmatrix}                                                       \frac{N_0}{P}
\end{pmatrix}^{N_R+N_SN_D}\nonumber\\
&\nonumber\\
+&\frac{N_SN_D}{\pi}\frac{(N_SN_D-1)!I_2}{(b\beta_1\sigma_{SD}^2)^{N_SN_D}}\frac{I_4}{(\frac{\beta_2\xi\sigma_{RD}^2}{4N_R})^{N_RN_D}}\nonumber\\
&\times\begin{pmatrix}                                                     \frac{N_0}{P}                                    \end{pmatrix}^{N_RN_D+N_SN_D}
.\end{align}
%%%%%%%%%%%%%%%%%%%%%%%%%%%%%%%%%%%%%%%%%%%%%%%%%%%%%%%%%%%%%%%%%%%%%%%%%ù
\section{Simulation and Results}
In this section, we present the results of our proposed system in terms of SER. The results are evaluated for several combinations of $N_s,N_R$ and $N_D$ using $QPSK$  modulation $(M=4)$, with a total power $P=20$ $db$, and $\sigma_{SR}^2=\sigma_{SD}^2=\sigma_{RD}^2=1$ and $N_0=1$.
In Fig 3, we present the analytical SER, Upper Bound simulation results with equal power and optimal power allocation for different number $(N_s,N_R,N_D)$ of antenna at the source relay and destination respectively. It can be seen from Fig 3 that the simulation results approaching the derived SER and approaching the upper bond at high SNR. Moreover, the system with optimal power allocation has low SER compared to the system with equal power.
%%%%%%%%%%%
\begin{figure}[htbp]
\centerline{\includegraphics[scale=0.6]{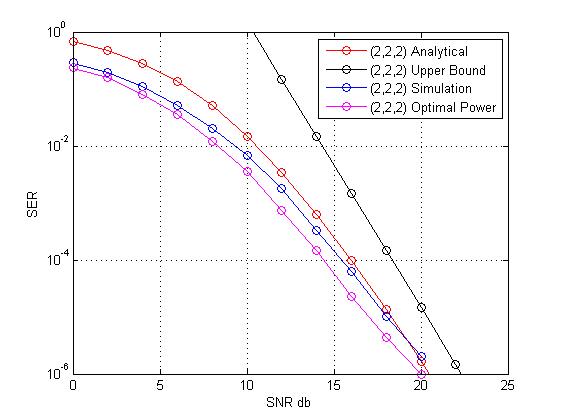}}
\caption{Analytical, Simulation, Upper Bound and Optimal Power allocation of SER for Strategies I.}
\label{fig3}
\end{figure}
%%%%%%%%%%%
In Fig 4, we plot the simulation results of SER for different diversity order and we get the following result:
\begin{itemize}
\item Increasing the number of antenna at the source and destination lead us to higher diversity order and better performance.
\item Fixed the number of antenna at the source and the destination and increase the number of antenna at the relay does not improve  specially at low SNR the performance since diversity order depend only on the number of antenna at the source and destination.
\end{itemize}
In Fig 5, we plot the analytical SER in both cases equal power and optimal power allocation for different diversity order. Furthermore, as expected increasing the number of antennas at the source and destination regardless the number of antennas at the relay improves the performance. Moreover, using optimal power enhances the SER compared with equal power.
In Fig 6, we plot the analytical and upper bound for the second strategy with same number of antennas at the relay and different number antennas at the destination and for different diversity order with equal power i.e. transmit power at the source is $P_s=P/2$ and at the each antenna of the relay is $P_{R,i}=P/(2N_r)$ where $1\leq i \leq N_R$, and as expected increasing the number of antennas at the relay, source or destination lead to better performance in terms of SER, also at high SNR the analytical approach to the upper bound.
%%%%%%%%%%%
\begin{figure}[htbp]
\centerline{\includegraphics[scale=0.6]{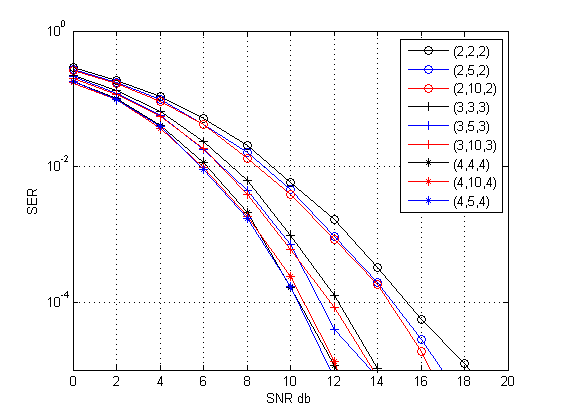}}
\caption{Comparison of Simulation SER for Different Diversity order for Strategies I.}
\label{fig4}
\end{figure}
%%%%%%%%%%%%%%%%%%%%%%%%%%%%%%%%%%
%%%%%%%%%%%%%%%%%%%%%%%%%%%%%%%%%%%%%%%%%%%%%%%%%%%%%%%%%%%%%%%%%%%%%%%%%%%%
\begin{table}[htbp]
\caption{OPTIMAL POWER ALLOCATION FOR STRATEGIES I}
\begin{center}
\begin{tabular}{|c|c|c|}
\hline
                \textbf{Antenna}& $\mathbf{P_s/P}$ & $\mathbf{P_R/P}$ \\
    \hline
    $\mathbf{N_S=1,N_R=1,N_D=1}$  & 0.6270 & 0.3730 \\
    \hline
    $\mathbf{N_S=2,N_R=2,N_D=2}$  & 0.8086 & 0.1914 \\
    \hline
    $\mathbf{N_S=3,N_R=3,N_D=3}$  & 0.9026 & 0.0974 \\
    \hline
\end{tabular}
\label{tab1}
\end{center}
\end{table}
\begin{figure}[htbp]
\centerline{\includegraphics[scale=0.6]{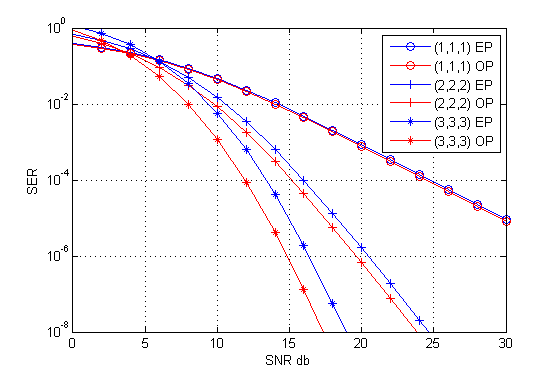}}
\caption{Analytical SER for Different Diversity order for Strategies I with equal power and optimal power allocation.}
\label{fig5}
\end{figure}
In Fig 7, we plot a comparative analysis of both strategies for the same diversity order for different number of antennas at the relay.

Let’s $R$ be the rate and depending on the number of antenna at the relay define as follow:
\begin{equation}\label{eq38}
R=\frac{\text{Number of transmitted symbol}}{\text{Number of all transmission}}	
\end{equation}	
In Table II, we summarize the number of antennas and the rate for different number of antennas as in \cite{b10}.
\begin{table}[htbp]
\caption{RATE FOR DIFFERENT NUMBER OF ANTENNA FOR STRATEGY II.}
\begin{center}
\begin{tabular}{|c|c|c|}
\hline
    $\mathbf{N_R}$& $\mathbf{N_D}$ & \textbf{Rate} \\
    \hline
    $2$  & 1 & 1/2 \\
    \hline
    $2$  & 2 & 1/2 \\
    \hline
    $3$  & 1 & 1/3 \\
    \hline
    $4$  & 1 & 1/3 \\
    \hline
\end{tabular}
\label{tab1}
\end{center}
\end{table}
\begin{figure}[htbp]
\centerline{\includegraphics[scale=0.6]{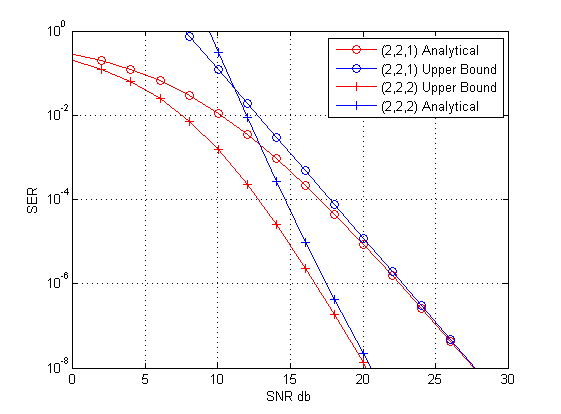}}
\caption{Analytical, upper Bound SER for Strategies II }
\label{fig6}
\end{figure}
It can be seen from Fig 7, that for the same diversity order both strategies have the same slope for the same diversity order but the second strategy has better SER performance compared to the first strategy when they have the same Rate. However, increasing the number of antennas at the relay lead to lower rate R and therefore bandwidth expansion.
\begin{figure}[htbp]
\centerline{\includegraphics[scale=0.6]{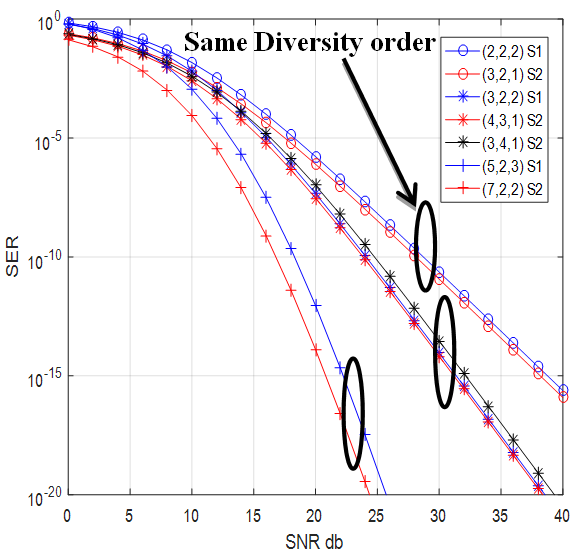}}
\caption{Comparison of Strategy I and strategy II for different diversity order }
\label{fig7}
\end{figure}

%%%%%%%%%%%%%%%%%%%%%%%%%%%%%%%%%%%%%%%%%%%%%%%%%%%%%%%%%%%%%%%%%%%%%%%%%
\section{CONCLUSION}
In this work, we analyze the performance of two strategies for single MIMO Relay. For the first strategy, we have derive the SER expression upper bound and optimal power allocation and the diversity order which are $N_s N_d+1$. The second strategy, which is based on transmit diversity at the relay using STBC has diversity order $N_sN_d+N_R$. When comparing the two strategies; we found that the first strategy has the advantage of low complexity of implementation compared to the second one. On the other hand, when both systems have the same diversity order and the same rate the second strategy performs better and has lower SER.Moreover, increasing the number of antennas at relay lead to lower rate compared to the first strategy which has always a rate of $1/2$. When we have two antennas at the relay we prefer the second strategy because it has the same rate and lower SER but the case when $N_R>2$, we prefer strategy I due to its high rate and lower complexity compared to strategy II .

%%%%%%%%%%%%%%%%%%%%%%%%%%%%%%%%%%%%%%%%%%%%%%%%%%%%%%%%%%%%%%%%%%%%%%
\section*{Appendix}
\subsection{Derivation of \eqref{19}}
\begin{align*}
   P_{S\rightarrow D}&= E_{g_{SD}}\{\Psi(\lambda_{SD})\}\\
   & = \int\limits_0^{\infty}\Psi(\lambda_{SD})f_{g_{SD}}(g)dg \\
   & = \frac{1}{\pi}\dfrac{N_{S}N_d}{\delta_{SD}^2}\int\limits_0^{\infty}\int\limits_0^{\left(\frac{M-1}{M}\right)\pi}\exp\left(-\left(\dfrac{bP_S}{N_0\sin^2\theta}+\frac{1}{\delta_{SD}^2}\right)g\right)\\
   &\left(1-\exp\left(\frac{-g}{\delta_{SD}^2}\right)\right)^{N_SN_D-1}d\theta dg
\end{align*}
Using the Binomial expansion $(1-x)^{N}=\sum\limits_{n=0}^{N_r-1}\displaystyle\binom{N}{n}(-1)^nx^n$, we get:
  \begin{align*}
    P_{S\rightarrow D} =& N_SN_d\sum\limits_{n=0}^{N_SN_d-1}\displaystyle\binom{N_SN_d-1}{n}(-1)^n \\
    &F\left(\dfrac{bP_S\delta_{SD}^2}{N_0\sin^2\theta}+n+1\right). 
  \end{align*}

%%%%%%%%%%%%%%%%%%%%%%%%%
\subsection{Derivation of \eqref{21}}
  \begin{align*}
P_{S\rightarrow D, R\rightarrow D}&= E_{f_{g_{SD}}}f_{g_{RD}}\{\Psi(\lambda)\}
\end{align*}
\begin{align*}
&=\frac{1}{\pi}\int\limits_0^{\left(\frac{M-1}{M}\right)\pi}\int\limits_0^{\infty}\Psi(\lambda_{SD})f_{g_{SD}}(g_{SD})dg_{SD}\\
&\int\limits_0^{\infty}\Psi( \lambda_{RD})f_{g_{RD}}(g_{RD})dg_{RD}d\theta\\
&=N_{S}N_d\sum\limits_{n=0}^{N_SN_d-1}\displaystyle\binom{N_SN_d-1}{n}(-1)^n\\
&F\left(\left(\dfrac{bP_S\delta_{SD}^2}{N_0\sin^2\theta}+n+1\right)\left(\dfrac{bP_R\delta_{RD}^2}{N_0\sin^2\theta}+1\right)\right)
\end{align*}

%%%%%%%%%%%%%%%%%%%%%%%%%%%%%%%%%%%%%%%%%%%%%%%%%%%%%%%%%%%%%%%%%%%%%%%%%%
\subsection*{Derivation of \eqref{22}}
\begin{align*}
P_e&\leq\dfrac{N_{S}N_dN_R}{\pi^2}\int\limits_0^{\left(\frac{M-1}{M}\right)\pi}\int\limits_0^{\left(\frac{M-1}{M}\right)\pi}\sum\limits_{n=0}^{N_SN_d-1}\sum\limits_{m=0}^{N_r-1}\displaystyle\binom{N_r-1}{n}\\
&\displaystyle\binom{N_SN_d-1}{m}\dfrac{(-1)^n}{\left(\dfrac{bP_S\delta_{SR}^2}{N_0sin^2\theta_1}+n\right)}\dfrac{(-1)^m}{\left(\dfrac{bP_S\delta_{SD}^2}{N_0\sin^2\theta_2}+m\right)}d\theta_1d\theta_2\\
&+\dfrac{N_{S}N_d}{\pi}\int\limits_0^{\left(\frac{M-1}{M}\right)\pi}\sum\limits_{\nu=0}^{N_SN_d-1}\displaystyle\binom{N_SN_d-1}{\nu}\dfrac{(-1)^\nu}{\left(\dfrac{bP_S\delta_{SD}^2}{N_0\sin^2\theta}+\nu\right)}\\
&\dfrac{1}{\left(\dfrac{bP_R\delta_{RD}^2}{N_0\sin^2\theta}+1\right)}d\theta
\end{align*}
Using the following result from \cite{b8} $\sum\limits_{\nu=0}^N\displaystyle\binom{N_r-1}{\nu}(-1)^\nu \frac{1}{x+\nu y}=\dfrac{N! y^N}{\Pi_{\nu=0}^{N}(x+\nu y)}$ and approximating at high SNR:
\begin{align*}
P_e&\leq\dfrac{N_{S}N_dN_R}{\pi^2}\dfrac{(N_r-1)!}{\left(\dfrac{bP_S\delta_{SR}^2}{N_0}\right)^{N_r}}\dfrac{(N_SN_d-1)!I_1I_2}{\left(\dfrac{bP_S\delta_{SD}^2}{N_0}\right)^{N_SN_d}}\\
&+\dfrac{N_{S}N_d}{\pi}\dfrac{(N_SN_d-1)!}{\left(\dfrac{bP_S\delta_{SD}^2}{N_0}\right)^{N_SN_d}}\dfrac{I_3}{\left(\dfrac{bP_R\delta_{RD}^2}{N_0}\right)}
\end{align*}
%%%%%%%%%%%%%%%%%%%%%%%%%%%%%%%%%%%%%%%%%%%%%%%%%%%%%%%%%%%%%%%%%%%%%%%
\subsection*{Derivation of \eqref{32}}
\begin{align*}
     P_{S\rightarrow D, R\rightarrow D}&=E_{f_{g_{SD}}}f_{g_{RD}}\{\Psi(\lambda)\} \\
     &=\frac{1}{\pi}\int\limits_0^{\left(\frac{M-1}{M}\right)\pi}\int\limits_0^{\infty}\Psi(\lambda_{SD})f_{g_{SD}}(g_{SD})dg_{SD}\\
     &\int\limits_0^{\infty}\Psi(\lambda_{RD})f_{g_{RD}}(g_{RD})dg_{RD}d\theta\\
     &\simeq\frac{1}{\pi}\int\limits_0^{\left(\frac{M-1}{M}\right)\pi}\int\limits_0^{\left(\frac{M-1}{M}\right)\pi}\int\limits_0^{\infty}\Psi_{\theta_1}(\lambda_{SD})f_{g_{SD}}(g_{SD})dg_{SD}\\
     &\int\limits_0^{\infty}\Psi_{\theta_2}(\lambda_{RD})f_{g_{RD}}(g_{RD})dg_{RD}d\theta_1 d\theta_2\\
     &=P_{S\rightarrow D}\times P_{R\rightarrow D}
     \end{align*}
     \begin{align*}
     &=N_{S}N_d\sum\limits_{n=0}^{N_SN_d-1}\displaystyle\binom{N_SN_d-1}{n}(-1)^nF\left(\dfrac{bP_S\delta_{SD}^2}{N_0\sin^2\theta}+n+1\right)\\
     \times &\sum\limits_{X_l\in \mathcal{C}, l\neq n}^{|C|}F\left(\Pi_{k=1}^{N_r}\left(\dfrac{P_r\lambda_{k,l}\delta_{rd}^2}{4N_0N_R\sin^2\theta}+1\right)^{N_d}\right)
  \end{align*}
  where 
  \begin{equation*}
    \Psi_{\theta_i}(\lambda)=\frac{1}{\pi}\int\limits_0^{\left(\frac{M-1}{M}\right)\pi}\exp\left(-\frac{b}{\sin^2\theta_i}\lambda\right)d\theta_i.
  \end{equation*}
%%%%%%%%%%%%%%%%%%%%%%%%%%%%%%%%%%%%%%%%%%%%%%%%%%%%%%%%%%%%%%%%%%%%%%%%%%%%%%%%%%%%%%%%%%%%%%%%%%%%%%%%%%%%%%%%%%%%%%%%%%%%%%%%%%%%%
%%%%%%%%%%%%%%%%%%%%%%%%%%%%%%%%%%%%%%%%%%%%%%%%%%%%%%%%%%%%%%%%%%%%%%%%%%%%%
%%%%%%%%%%%%%%%%%%%%%%%%%%%%%%%%%%%%%%%%%%%%%%%%%%%%%%%%%%%%%%%%%%%%%%%%%%%%%
%%%%%%%%%%%%%%%%%%%%%%%%%%%%%%%%%%%%%%%%%%%%%%%%%%%%%%%%%%%%%%%%%%%%%%%%%%%%%
%%%%%%%%%%%%%%%%%%%%%%%%%%%%%%%%%%%%%%%%%%%%%%%%%%%%%%%%%%%%%%%%%%%%%%%%%%%%%
%%%%%%%%%%%%%%%%%%%%%%%%%%%%%%%%%%%%%%%%%%%%%%%%%%%%%%%%%%%%%%%%%%%%%%%%%%%%%
%%%%%%%%%%%%%%%%%%%%%%%%%%%%%%%%%%%%%%%%%%%%%%%%%%%%%%%%%%%%%%%%%%%%%%%%%%%%%
%%%%%%%%%%%%%%%%%%%%%%%%%%%%%%%%%%%%%%%%%%%%%%%%%%%%%%%%%%%%%%%%%%%%%%%%%%%%%
%\section*{References}

\end{document}